\documentclass{article}
\newcommand{\bfr}{\begin{flushright}}
\newcommand{\efr}{\end{flushright}}
 
\begin{document}
\title{Thermodynamic potential for compactified bosonic strings
}
\author{Kiyoshi Shiraishi\\
Department of Physics, Tokyo Metropolitan University\\
Setagaya,
Tokyo, 158 Japan
}
\date{Il Nuovo Cimento {\bf A100} (1988) 683--692}
\maketitle
\begin{abstract}
We discuss the inclusion of chemical potentials of the Kaluza-Klein
charges in the partition function of the bosonic string with a
compactified dimension on a circle. The construction of the
thermodynamic potential is achieved by the path integral method at
one-loop level. Duality symmetry in the dependence on compactification
scale is examined. A modular-invariant expression for the thermodynamic
potential is also presented.
\end{abstract}

\bigskip

String theories have a particle content with unlimitedly increasing
level density. Thermodynamics of such system has been investigated by
many people \cite{1,2}, in the context of the very early universe.

 On the other hand, the calculation of thermo-partition function for
bosonic strings is given by Polchinski \cite{3}. He used the path
integral method with compact time direction. McClain and Roth \cite{4}
and O'Brien and Tan \cite{5} showed modular invariance of the partition
function explicitly. The path integral method is useful for
clarification of the symmetry in the partition function at finite
temperature as well as in the case with compactification.

On compactified space, gauge symmetry is induced not only in field
theory but also in string theories. In string theories, moreover,
larger symmetry than isometry of compact manifold can emerge. The gauge
bosons come from the winding soliton state in addition to the
Kaluza-Klein $U(1)$ gauge bosons and they can enlarge the symmetry \`a
la Frenkel and Kac \cite{6}.

The appearance of gauge fields is indispensable for the search for
lower-dimensional string theories. Recently it is found that
compactification of closed superstrings in rather complicated way
brings about the realistic gauge group \cite{7}. Perhaps,
compactification and gauge symmetry may affect each other through
evolution of hot early universe.

In thermodynamics, only chemical potentials that correspond to mutually
commuting charges can be introduced for a non-Abelian symmetry
\cite{8}. The Kaluza-Klein $U(1)$ bosons, which correspond to the
Cartan subalgebra of the gauge group, are associated with these
charges. There is a possibility that finite density of charges may
bring a crucial key to the early universe, in the context of
``superstring cosmology'' \cite{1,2}.

In this paper, we construct the thermodynamic potential which contains
the chemical potentials for $U(1)$ charges by the path integral
treatment of Polyakov's string at one-loop level. For simplicity, we
consider closed bosonic strings compactified on one-dimensional torus
here.

We start with the action for the first-quantized bosonic string
written as 
\begin{equation}
S=\frac{T}{2}\int
d^2\sigma\left(\sqrt{g}g^{ab}\partial_aX^M\partial_bX^N
G_{MN}+\varepsilon^{ab}\partial_aX^M\partial_bX^N
B_{MN}\right)\,,
\label{eq1}
\end{equation}
where $X^M$ ($M = 0,\dots,25$) are bosonic fields, $G_{MN}$ is the
background geometry and $B_{MN}$ is the background antisymmetric tensor
field. $T$ is the string tension. 

We can take a world-sheet metric
$g_{ab}$, without loss of generality, as follows: 
\begin{equation}
d^2\sigma=g_{ab}d\sigma^ad\sigma^b=\exp[\varphi]|d\sigma_1+\tau
d\sigma_2|^2 \quad(0\le\sigma_1\le1,~ 0\le\sigma_2\le1 )\,.
\end{equation}

Then the action (\ref{eq1}) can be rewritten as
\begin{eqnarray}
S&=&\frac{T}{2}\int d^2\sigma \,\tau_2\cdot\nonumber \\
& &\quad\left(G_{MN}\left\{\partial_1X^M\partial_1X^N+
\frac{1}{\tau_2}(\partial_2-\tau_1\partial_1)X^M
(\partial_2-\tau_1\partial_1)X^N\right\}\right.\nonumber \\
&
&\qquad\left.+\frac{2i}{\tau_2}B_{MN}\partial_1X^M\partial_2X^N\right)\,,
\end{eqnarray}
where modular parameter $\tau=\tau_1+i\tau_2$.

Note that this action is not expressed as light-cone form. We have
omitted possible ghost and other terms which have no concern with the
following derivation.

Let us consider compactification on a torus. We regard the twenty-fifth
dimension as a circle. It is well known that, if the radius of the
circle satisfies the Frenkel-Kac condition, $SU(2)\times SU(2)$ gauge
symmetry is realized. Apart from the winding sector, we have two $U(1)$
symmetries, that is, there are two Kaluza-Klein gauge fields.

There is no surprise about the fact in string theories because the
second ``Kaluza-Klein gauge boson'' is contained in the antisymmetric
tensor field $B_{MN}$. 

Now we introduce two chemical potentials
associated with Kaluza-Klein charges into the partition function as a
path integral form. First, we give the background metric of target
space as follows:
\begin{equation}
\left\{\begin{array}{ll}
G_{00}= 1 + A_0^2 & \\
G_{ij}=\delta_{ij} & (i, j=1,\dots , 24)\\
G_{0I}=G_{I0}= A_0 & (I=25)\\
\mbox{and} & \\
G_{II}= 1, \quad\mbox{otherwise}\quad G_{MN}=0
\end{array}
\right.\,.
\end{equation}

These are just zero-mode ans\"atze in Kaluza-Klein theory, except for
only the zeroth component of the vector boson takes a nonzero value.

Second, the background antisymmetric tensor field should take the
following form:
\begin{equation}
B_{0I}=-B_{I0}=-B_{0}\,, \quad\mbox{otherwise}\quad B_{MN}=0\,.
\end{equation}

Next we should note that $G_{MN}$ and $B_{MN}$ will only have influence
on string zero modes \cite{9}. For the first time, we consider that
$X^0$ is periodic in $\sigma_2$ modulo $\beta$ \cite{3}. $\beta$ will be
interpreted as the inverse of temperature. Furthermore, $Y\equiv X^I$ is
periodic in $\sigma_1$ and $\sigma_2$ modulo
$2\pi r$, where $r$ is the radius of the torus. Thus, we write
zero-mode pieces of $X^0$ and $Y$ as 
\begin{equation}
\bar{X}^0=\beta n\sigma_2\quad\mbox{and}\quad\bar{Y}=2\pi r (m\sigma_1+
\ell \sigma_2)\,,
\label{eq6}
\end{equation}
where $n$, $m$, and $\ell$ are integers.

One then obtains for the action by use of all the above ans\"atze
\begin{equation}
S=S_p+\bar{S}\,,
\end{equation}
where $S_p$ is the action expressed in terms of periodic pieces of
string coordinates and
\begin{eqnarray}
\bar{S}&=&\frac{T}{2}\left[\frac{\beta^2n^2}{\tau_2}+2iB_0(2\pi r)
\beta nm+\tau_2
(2\pi r)^2m^2\right.\nonumber \\
& &\qquad\left.+\frac{1}{\tau_2}((2\pi r)(\ell-\tau_1m)+A_0\beta n)^2
\right]\,.
\end{eqnarray}

The path integral is proportional to
\begin{equation}
\sum_{n, m, \ell=-\infty}^\infty\exp[-\bar{S}]\,.
\end{equation}

The summation over $\ell$ can be transformed into the familiar form by
utilization of the inversion relation of the theta function \cite{10}.
Then we find 
\begin{eqnarray}
& &\sum_{\ell}\exp\left[-\frac{T}{2\tau_2}((2\pi r)(\ell-
\tau_1m)+A_0\beta n)^2\right]\nonumber \\ &=&
\sqrt{\frac{2\pi\tau_2}{T(2\pi r)^2}}\vartheta_3
\left(\tau_1m-\frac{A_0\beta n}{2\pi r}\Big|\frac{2\pi
i\tau_2}{T(2\pi r)^2}\right)\nonumber \\ &=&
\sqrt{\frac{2\pi\tau_2}{T(2\pi r)^2}}\sum_\ell
\exp\left[-\frac{\pi\tau_2}{2\pi Tr^2}\ell^2+2\pi
i\left(\tau_1m-\frac{A_0\beta n}{2\pi r}\right)\ell\right]\,.
\end{eqnarray}
After all, we have
\begin{eqnarray}
& &\sum_{n, m, \ell}\exp\left[-\bar{S}\right]\nonumber \\ &=&
\sqrt{\frac{\tau_2}{2\pi Tr^2}}\sum_{n, m, \ell}
\exp\left[-\pi\tau_2\left\{(2\pi
Tr^2)^{1/2}m^2+\frac{\ell^2}{2\pi Tr^2}\right\}+ 2\pi
i\tau_1m\ell\right.\nonumber \\
& &\left.-\frac{\beta^2n^2T}{2\tau_2}-i\beta n(2\pi T)^{1/2}
\left\{B_0(2\pi T)^{1/2}rm+\frac{A_0}{(2\pi
T)^{1/2}r}\right\}\ell\right]\,.
\end{eqnarray}

Among these sums, the term $n=0$ (with sum over $m$ and $\ell$)
reproduces the ``correction factor'' $F_2$ in ref.~\cite{11} with $a =
(2\pi T)^{1/2}r$. The $n=0$ part gives vacuum energy or cosmological
constant, while the rest is connected with thermo-partition function.

Now we can write down the thermodynamic potential for bosonic strings.
We should follow the prescription to incorporate chemical potentials;
for instance, it is found in ref.~\cite{12}. Chemical potentials are
indentified with ``imaginary'' electric potentials. In our case, we
will set
\begin{equation}
A_0\rightarrow -i\mu_1\quad\mbox{and}\quad B_0\rightarrow -i\mu_2.
\end{equation}
We do not pay attention to normalization here.

The path integral for the free energy is then reduced to
\begin{eqnarray}
F&=&\int_0^\infty\frac{d\tau_2}{2\pi\tau_2^2}\int_{-1/2}^{1/2}d\tau_1
\exp
[4\pi\tau_2] (2\pi\tau_2)^{-1/2}
|f(\exp[2\pi i\tau])|^{-48}\nonumber \\
& &\cdot a\sqrt{\tau_2}\sum_{n=1}^\infty\sum_{m, \ell}
\exp[-\pi\tau_2(a^{-2}m^2+a^2\ell^2)+2\pi i\tau_1m\ell]\nonumber \\
& &\cdot\exp\left[-\frac{\beta^2n^2T}{2\tau_2}\right]\cosh\{\beta n 
(2\pi T)^{1/2}(\mu_2a^{-1}m+\mu_1a\ell)\}\,,
\label{eq13}
\end{eqnarray}
where the notation of factors which come from oscillator pieces is the
same as ref.~\cite{3}. Over all volume factor is already discarded.

It is easy to compare (\ref{eq13}) with the thermodynamic potential
calculated from the particle spectrum of bosonic strings. We note the
mass spectrum:
\begin{equation}
M^2=4\pi T\left(N+\tilde{N}-2+\frac{1}{2}(a^{-2}m^2+a^2\ell^2)\right)\,,
\label{eq14}
\end{equation}
with the constraint
\begin{equation}
N=\tilde{N}-m\ell\,,
\label{eq15}
\end{equation}
where $N$ and $\tilde{N}$ are the occupation numbers of oscillators.

We can obtain the expression (\ref{eq13}) by summing the thermodynamic
potential of one particle over the spectrum (\ref{eq14}) with constraint
(\ref{eq15}). Incidentally, the thermodynamic potential in Kaluza-Klein
theory is shown in ref.~\cite{13} with inclusion of the sum over a
single label $\ell$.

Of course, the expression (\ref{eq13}) is a formal one, because it is
divergent due to the presence of a tachyonic mode. If we want to
calculate a finite result at present level, we need such a prescription
as presented by Sakai and Senda \cite{14}; that is,
\begin{equation}
|f(\exp[2\pi i\tau])|^{-48}\rightarrow|f(\exp[2\pi i\tau])|^{-48}-1\,.
\label{eq16}
\end{equation}

By the way, suppose that the system is neutral, i.e. $\mu_1=\mu_2=0$.
Duality relation, which is demonstrated by many authors \cite{15} on
compactification on tori at zero temperature, is preserved even at
finite temperature. The free energy is invariant under the duality
transformation $a\leftrightarrow a^{-1}$.

At finite density, the thermodynamic potential may not be generally
invariant. If we assume no further normalization of two chemical
potentials and that $\mu_1$ and $\mu_2$ are constant, the duality
symmetry holds only when $\mu_1=\mu_2$. Needless to say, actually,
$\mu$'s are possibly function of $r$ as well as $\beta$ in physical
applications, for conservation of charges in total system.
Nevertheless, although the detailed dependence on $r$ and $\beta$ may be
complicated, it is rather natural to factorize the $r$-dependence of
chemical potentials by using $\mu_1$ and $\mu_2$ because charges depend
on the radius $r$ as the case of Kaluza-Klein theory. Therefore it can
happen that the duality symmetry is broken at finite density of
charges. The thought along this way probably provides a solution to
instability reported by Matsuo \cite{2}.

There is a difficulty in the physical application in fact. The problem
is concerned with the existence of massless charged bosons. If $a=1$,
i.e. $2\pi Tr^2=1$, exact $SU(2)\times SU(2)$ gauge symmetry is
realized. Corresponding to $|m|$, $|\ell|=1$, four massless
``W-bosons'' emerge at this ``Frenkel-Kac point''. 

Massless charged
bosons always condensate at any temperature \cite{16}. Thus, these
massless particles must be treated separately in an application. When
$a\ne 1$, chemical potentials can take nonzero values in the
thermodynamic potential for massive W-bosons. This case may occur in
the evolution of early universe. We must study it in detail under
cosmological consideration.

Another direction of application is a derivation of the ``colour''
singlet partition function \cite{17}. This subject will be studied
elsewhere.

Finally we examine the modular invariance of the thermodynamic
potential. The modular invariance is not manifest in expression
(\ref{eq13}). We will show the modular invariance of the thermodynamic
potential by means of almost the same way as shown in ref.~\cite{4,5}.

At first we change the zero-mode ansatz (\ref{eq6}) on $X^0$ as
\begin{equation}
\bar{X}^0=\beta(n_1\sigma_1+n_2\sigma_2)\,,
\end{equation}
where $n_1$ and $n_2$ are integers.

Now $\bar{S}$ is modified to be
\begin{eqnarray}
\bar{S}&=&\frac{T}{2}\tau_2\left((1+A_0^2)\beta^2\left\{
n_1^2+\frac{1}{\tau_2}(n_2-\tau_1n_1)^2
\right\}\right.\nonumber \\
& &+2A_0\beta(2\pi
r)\left\{n_1m+\frac{1}{\tau_2}(n_2-\tau_1n_1)(\ell-\tau_1m)\right\}\nonumber
\\
& &\left.+(2\pi
r)^2\left\{m^2+\frac{1}{\tau_2}(\ell-\tau_1m)^2\right\}-
\frac{2i}{\tau_2}B_0\beta(2\pi r)(n_1\ell-n_2m)\right)\,.
\label{eq18}
\end{eqnarray}
We then obtain the following new expression of the thermodynamic
potential:
\begin{eqnarray}
F&=&\int_F\frac{d^2\tau}{4\pi\tau_2^2}\exp[4\pi\tau_2](2\pi\tau_2)^{-12}
|f(\exp[2\pi i\tau])|^{-48}\nonumber \\
&\cdot&{\sum_{n_1n_2}}'\sum_{m,\ell}\exp\left[
-\frac{T}{2}\tau_2\left((1+A_0^2)\beta^2\left\{
n_1^2+\frac{1}{\tau_2}(n_2-\tau_1n_1)^2
\right\}\right.\right.\nonumber \\
& &\qquad+2A_0\beta(2\pi
r)\left\{n_1m+\frac{1}{\tau_2}(n_2-\tau_1n_1)(\ell-\tau_1m)\right\}\nonumber
\\
& &\left.+(2\pi
r)^2\left\{m^2+\frac{1}{\tau_2}(\ell-\tau_1m)^2\right\}-
\frac{2i}{\tau_2}B_0\beta(2\pi r)(n_1\ell-n_2m)\right)\,,
\end{eqnarray}
where the prime denotes the omission of $n_1=n_2=0$ term.

Note that the integration region of $\tau$ is restricted in the
fundamental region $F$, in which $\tau$ satisfies $-1/2< \tau_1<1/2$,
$\tau_2>0$ and $|\tau|>1$. This expression clearly exhibits manifest
modular invariance. One can easily check the invariance under
transformations $\tau\rightarrow\tau+1$ and $\tau\rightarrow-1/\tau$.

Now, we must prove the equivalence of the expressions (\ref{eq13}) and
(\ref{eq18}) for the thermodynamic potential. The procedure for this
purpose is as follows. Consider particular values in $n_1$ and $n_2$. We
denote the greatest common divisor of $n_1$ and $n_2$ by $p$. Then we
can write
$n_1=pc$ and $n_2= pd$. $c$ and $d$ are relative primes. (If $n_1$ or
$n_2=0$, take them as $(n_1, n_2)=(0, p)$ or $(p, 0)$, i.e. $(c, d)= (0,
1)$ or $(1, 0)$.) 

Change variable $\tau\rightarrow\tau'$ as
\begin{equation}
\tau_2'=\frac{\tau_2}{|\tau|^2c^2+d^2-2cd\tau_1}\,.
\label{eq20}
\end{equation}
This transformation can be regarded as modular transformation 
$\tau'=(a\tau+b)/
(-c\tau+d)$ \cite{4,5}. There remains the arbitrariness of the choice of
$a$ and $b$. The same transformation reduces the other terms to the
following form:
\begin{eqnarray}
& &\tau_2cm+\frac{1}{\tau_2}(d-\tau_1c)(\ell-\tau_1m)=
\frac{1}{\tau_2'}(\ell'-\tau_1'm')\nonumber \\
& &\mbox{and}\nonumber \\& &
\tau_2m^2+\frac{1}{\tau_2}(\ell-\tau_1m)^2=
\tau_2'm'^2+\frac{1}{\tau_2'}(\ell'-\tau_1'm')^2\,,
\end{eqnarray}
where $m'=dm-c\ell$ and $\ell'=bm+a\ell$.

The parameters $a$ and $b$ only appear in the labels of the sums. It has
been proved that the modular transformation which maps the fundamental
region into the strip region in which $\tau$ satisfies $-1/2<\tau_1<
1/2$ and $\tau_2>0$ exists, and is unique, for given fixed $c$ and $d$
\cite{4,5}. Thus, we take such values for $a$ and $b$. $a$ and $b$ are
also relative primes, for they satisfy $ad+bc=1$. We can then say
that
$m'$ and
$\ell'$ run over all integers as $m$ and $\ell$ take integer values.

The integer $p$ which is factored out in (\ref{eq20}) plays the same
role of label as $n$ in (\ref{eq13}). Now it is found that the
integration in $F$ connected with the various combinations $(c,d)$,
with sums over
$m$, $\ell$ and $p$, makes up the integral expression (\ref{eq13}) in
the strip region (of course up to the inversion relation). Then the
equivalence of (13) and (20) is established. After all we can show the
modular invariance of the thermodynamic potential.

In summary, we derived the expression of the thermodynamic potential
for compactified bosonic strings from Polyakov path integral and
discussed its duality relation and modular invariance.

We must comment on the application to cosmology. The divergence due to
tachyon can be removed by means of the prescription (\ref{eq16}), but
it spoils the modular invariance. Of course, the divergence of the
potential is absent when we consider superstrings \cite{18} including
heterotic strings \cite{19}. We must extend the result in this paper to
the supersymmetric case. It will be useful to employ the Green-Schwarz
action \cite{20,21} at the starting point. Further analysis on the
construction will be reported elsewhere.

 Applications to heterotic strings make the cosmological model
realistic for the large gauge group. At the same time, gauge symmetry
breaking becomes an important issue in cosmological evolution. Recently
various compactification schemes are examined for heterotic \cite{9,22}
as well as closed superstrings \cite{7}. It is interesting to
investigate a deep connection between thermodynamics and
compactifications in the early universe. The mechanisms generating
and/or breaking gauge symmetry by background fields are necessary for
constructing a realistic gauge group. The finite temperature and
density effect on these mechanisms \cite{23} will be a substantial
subject for theoretical research of evolutional universe. 

We must check
the efficiency of the canonical description and thermal and chemical
equilibrium in physical applications. Rather speculatively speaking,
the achievement of the string field theory will lead to the
microcanonical description for thermodynamics of strings.

\bigskip

I would like to thank S. Saito for reading this manuscript. I would
also like to thank Iwanami F\=ujukai for financial aid.


\end{document}